\documentclass[11pt]{article}
\usepackage{geometry}               
\usepackage{graphicx}
\usepackage{amssymb}
\usepackage{epstopdf}
\usepackage{url}
\newcommand{\be}{\begin{equation}}
\newcommand{\bea}{\begin{eqnarray}}
\newcommand{\beq}[1]{\begin{equation}\label{#1}}
\newcommand{\eq}[1]{Eq.~(\ref{#1})}

\newcommand{\ee}{\end{equation}}
\newcommand{\eea}{\end{eqnarray}}
\newcommand{\eeq}{\end{equation}}
\newcommand{\lsim}{\!\mathrel{\hbox{\rlap{\lower.55ex \hbox{$\sim$}} \kern-.34em \raise.4ex \hbox{$<$}}}}
\newcommand{\gsim}{\!\mathrel{\hbox{\rlap{\lower.55ex \hbox{$\sim$}} \kern-.34em \raise.4ex \hbox{$>$}}}}

\begin{document} 
 
\begin{titlepage}
\flushright{MCTP-07-11\\}
\vspace{1in}
\begin{center}
{\Large \bf Natural Dark Matter from an Unnatural Higgs Boson}
\vspace{.07in}
{\Large \bf and New Colored Particles at the TeV Scale}

\vspace{0.5in}
{\bf Aaron Pierce$^{1}$, Jesse Thaler$^{2,3}$}

\vspace{.5cm}

{\it $^{1}$ Randall Laboratory, Physics Department, University of Michigan, \\ Ann Arbor, MI 48109}

\vspace{0.2cm}

{\it $^{2}$ Department of Physics, University of California, Berkeley, CA 94720
}

\vspace{0.2cm}

{\it $^{3}$ Theoretical Physics Group, Lawrence Berkeley National Laboratory, \\ Berkeley, CA 94720}
\end{center}
\vspace{0.8cm}
\begin{abstract}
The thermal relic abundance of Dark Matter motivates the existence of new electroweak scale particles, independent of naturalness considerations.  However, most unnatural Dark Matter models do not ensure the presence of new particles charged under $SU(3)_C$, resulting in challenging LHC phenomenology.  Here, we present a class of models with scalar electroweak doublet Dark Matter that require a host of colored particles at the TeV scale.  In these models, the Higgs boson is apparently fine-tuned, but the Dark Matter doublet is kept light without any additional fine-tuning.  
\end{abstract}
\end{titlepage}

\section{Motivation}

The discovery that the expansion of the universe is accelerating, coupled with the rise of the landscape
picture in string theory, has led to a questioning of naturalness as a motivation for new physics
at the weak scale \cite{NimaSavas}.  After all, the fine-tuning associated with the cosmological constant (CC) problem is far greater than the fine-tuning of the (mass)$^2$ parameter of the Higgs boson.  If the solution to these two apparent tunings are intertwined, one should account for the more severe tuning for the CC before making statements about naturalness and particle physics at the weak scale.  For example, it is possible that once the CC is forced to be small, the Higgs boson mass might appear fine-tuned.  

Aside from naturalness considerations, the strongest motivation for new physics at the weak scale is the presence of Dark Matter in our universe.   Using the observed Dark Matter density, a typical thermal relic abundance calculation points to a thermally averaged cross section of \cite{KandT}:
\begin{equation}
\langle \sigma v \rangle = 0.1 \mbox{ pb}.
\end{equation}
When translated to a mass scale, $\langle \sigma v \rangle = \alpha / m^{2}$, we find $m \approx 100$ GeV.  Thus, one expects a weakly-interacting Dark Matter particle near the weak scale.  

But considerations of Dark Matter alone do not promise a rich TeV-scale collider phenomenology.  As enunciated clearly in \cite{StrumiaDM}, the addition of a multiplet charged under $SU(2)_{L}$ could minimally account for the Dark Matter, but might not be kinematically accessible at the Large Hadron Collider (LHC).  Similarly, the Inert Doublet Model \cite{Lawrence, Inert} shows that even if the Dark Matter is kinematically accessible at the LHC, observation may be difficult after including the Standard Model backgrounds.  A similar expectation is obtained for the case where an $SU(2)_L$ singlet coupled to the Higgs boson makes up the Dark Matter, as in \cite{NewMinimal,BurgessMcDonald}.  

On the other hand, spectacular LHC signatures will exist if new colored particles occur in association with the Dark Matter particles.  Most {\it natural} models---weak-scale supersymmetry (SUSY), composite Higgs models, little Higgs theories, technicolor models, and universal extra dimensions---do have TeV-mass colored particles to regulate the top-loop contribution to the Higgs potential.  In SUSY, the gluinos and squarks provide the excitement at the LHC, though Dark Matter is due to neutralinos.  It is the overall structure of supersymmetric models---motivated by naturalness considerations---that ensures its discovery at the LHC, not the existence of Dark Matter alone.  

Should we expect new TeV-scale colored particles in theories where the Higgs (mass)$^2$ parameter is \emph{unnatural}?  While the answer is no in the minimal models enumerated above, it is worthwhile to consider various motivations for new colored states in particle models of Dark Matter.  These models are of the most immediate interest---colored particles may be visible within the first $10 \mbox{ fb}^{-1}$ of data at the LHC. This paper seeks to answer the question: are there motivated structures that ensure the presence of TeV-scale colored particles even when the Higgs boson (mass)$^{2}$ parameter is fine-tuned?

One such example already exists in the the literature: Split SUSY \cite{NimaSavas,Giudice}. There, the scalar superpartners of the Standard Model are taken to be ultra-heavy.  The model is apparently fine-tuned, but the correct Dark Matter abundance can be recovered \cite{Giudice,Pierce} by placing the fermionic partners near the weak scale.  Since the masses of these fermionic partners are protected by chiral symmetries ($R$ symmetries and a Peccei-Quinn symmetry), it is only a ``logarithmic fine-tuning'' that the masses of these states coincide with the weak scale, and therefore technically natural.  These fermionic partners can lead to interesting collider signatures.  In particular, the presence of long-lived gluinos make this scenario an optimistic one for the LHC \cite{SplitCollider}.  

This paper is complementary, and fills a gap in unnatural model building. Even with a fine-tuned Higgs mass, the existence of new colored particles can be guaranteed by insisting that the mass of a scalar Dark Matter particle requires no additional quadratic fine-tuning.  One might ask: if one is unconcerned by the fine-tuning of the Higgs (mass)$^{2}$, why motivate new physics on the basis of the fine-tuning of a Dark Matter particle?  

There are two valid responses to this question.  One answer emerges in the context of multi-verse theories.   There are potentially strong environmental selection pressures on two dimensionful parameters: the CC and the Higgs boson mass \cite{Friendly}.  All other things equal, unless the CC and the Higgs mass take on values close to their fine-tuned values, the universe would be devoid of structure \cite{WeinbergCC} and no atoms beyond helium would exist \cite{Donoghue,NimaSavas}, two truly dramatic outcomes.  While the presence of Dark Matter certainly affects the details of structure formation \cite{Tegmark:2005dy}, it is less clear that a sharp catastrophe occurs in the absence of a tuning for the Dark Matter mass. Thus, dynamics should ensure its lightness, and if it is a scalar, its mass must be protected.  

Alternatively, one might remain agnostic and simply explore this novel structure and its phenomenology in preparation for upcoming experiments at the weak scale.  The theory explored here is one example of a novel class of models where the Higgs boson and the Dark Matter are intimately connected.  It is an attractive possibility that the two weak scale particles that we know the least about might be so closely related.

The idea is to relate the lightness of the Higgs boson to the lightness of the Dark Matter particle via a symmetry.  Even though both masses are quadratically sensitive to the cut-off, if the symmetry were perfect, the Dark Matter (mass)$^{2}$ and the Higgs boson (mass)$^{2}$ would be identical.  Thus, a single fine-tuning can accomplish the task of placing both particles at the weak scale.  The requirement that there be a small splitting between these particles (and that only the Higgs field acquires a vacuum expectation value), tells us that this symmetry must be softly broken.  

Indeed, the requirement that the scalar Dark Matter particle be at the weak scale tells us where the symmetry must be broken: within a loop factor of the weak scale and thus within kinematic reach of the LHC.  Models of this type require the Standard Model Yukawa couplings to be symmetrized, and hence new quark partners.  The result is a prediction of new colored states at the TeV scale and an interesting collider phenomenology.  In natural theories, the symmetry that protects the Higgs boson mass imply a proliferation of Standard Model partners; in these models the symmetry that protects the Higgs/Dark Matter splitting leads to a similar proliferation.  In this way, weakly-interacting Dark Matter can serve as a motivation for strongly-interacting new physics at the LHC.

In the next section, we give a simple model of technically natural scalar Dark Matter.  Its Dark Matter properties are analyzed in Sec.~\ref{sec:dm}, and its collider implications in Sec.~\ref{sec:collider}.  Sec.~\ref{sec:othersym} explores possible generalizations, and we conclude with a discussion of Dark Matter and the landscape.

\section{Natural Dark Matter from an Unnatural Higgs}
\label{sec:themodel}

To realize technically natural scalar Dark Matter, we put Dark Matter and the Higgs in the same multiplet of some enhanced symmetry $G$.   The mass splitting between the Higgs boson and Dark Matter will be controlled by the mass scale of $G$ symmetry breaking.   Because we do not want an extra fine-tuning to control the scale of $G$ symmetry breaking, we assume that $G$ is ultimately broken by some form of dimensional transmutation.  

The simplest example of an enhanced symmetry $G$ is an exchange symmetry $\mathbf{S}_2$.  Under this symmetry, the Higgs doublet $h$ is mapped to a Dark Matter doublet $\phi$, 
\beq{eq:exchange}
\mathbf{S}_2 \colon \quad  h \leftrightarrow \phi.
\ee
To ensure that at least one component of $\phi$ is exactly stable, we will also enforce a $\mathbf{Z}_2$ symmetry,
\be
\mathbf{Z}_2 \colon \quad \phi \rightarrow -\phi.
\ee
The $\mathbf{S}_2$ and $\mathbf{Z}_2$ symmetries do not commute, so $h$ and $\phi$ transform under a two dimensional representation of the dihedral group $\mathbf{D}_4$, which can arise in orbifold constructions.  In the Higgs/Dark Matter sector, these symmetries permit the following interactions:
\be
\label{eqn:scalarpotential}
V(h,\phi) = m^2\left(|h|^2 + |\phi|^2\right) + \lambda_{1} \left(|h|^4 + |\phi|^4\right) + \lambda_{3} |h|^2 |\phi|^2 + \lambda_{4} |h\phi^\dagger|^2 + \lambda_{5} \mathop{\rm Re} \left( h^\dagger \phi h^\dagger \phi \right). 
\ee
This is the $\mathbf{S}_2$ symmetric form of the more general Inert Doublet potential \cite{Lawrence}. The  $\mathbf{S}_2$ symmetry sets the $|\phi|^4$ coupling equal to the $|h|^{4}$ coupling.  The Dark Matter properties of the Inert Doublet model have been recently qualitatively discussed in \cite{Lawrence} and explored in some in detail in \cite{Inert}.  In Sec.~\ref{sec:relic}, we review the thermal relic abundance calculation for $\phi$ with this potential; this will pinpoint the region of most phenomenological interest for this model at the LHC.

To couple the Higgs to Standard Model fermions, the $\mathbf{S}_{2}$ symmetry will force us to introduce partners of either the fermion  $SU(2)_{L}$-doublets or  $SU(2)_{L}$-singlets or both.  This phenomenon is not confined to the particular model considered here.  The combination of the $\mathbf{Z}_2$ that keeps the $\phi$ field stable and the $G$  that relates the Higgs field to the $\phi$ will quite generically force the inclusion of partner particles. As an illustration, consider the case with partners for both types of fermions.  Then, the $\mathbf{S}_2$ symmetry from \eq{eq:exchange} is actually expanded to an $\mathbf{S}^A_2 \times  \mathbf{S}^B_2$ symmetry.\footnote{The semi-direct product $\mathbf{Z}_2 \times \mathbf{S}^A_2 \times \mathbf{S}^B_2$ is a finite group of order 32.}  Under these symmetries, the Standard Model doublets $q$ and $\ell$ and Standard Model singlets $u^c$, $d^c$, and $e^c$ for each generation are mapped to partner particles: $Q$, $L$, $U^c$, $D^c$, and $E^c$. 
\beq{eqn:xchange1}
\mathbf{S}^A_2 \colon \quad  h \rightarrow \phi, \quad q \leftrightarrow Q, \quad \ell \leftrightarrow L.
\ee
\beq{eqn:xchange2}
\mathbf{S}^B_2 \colon \quad  h \rightarrow \phi, \quad u^c \leftrightarrow U^c, \quad d^c \leftrightarrow D^c, \quad e^c \leftrightarrow E^c.
\ee
We assume that the $\mathbf{S}_2$ is soften broken when partner particles get vector-like masses.  This requires the introduction of conjugate fields: $Q^c$, $L^c$, $U$, $D$, and $E$.  All of these partners are odd under the $\mathbf{Z}_2$ symmetry.

The only Yukawa couplings compatible with the symmetries are (we write only the up quark sector couplings for simplicity):
\beq{eqn:yukawa}
\mathcal{L}_{\rm Yukawa} = \lambda_u \left(q h u^c + Q \phi u^c + q \phi U^c + Q h U^c \right).
\ee
The $\mathbf{S}^A_2 \times  \mathbf{S}^B_2$ symmetry ensures minimal flavor violating Yukawa couplings, \emph{i.e.}\ no new directions in flavor space are introduced via the Yukawa couplings.  While the $\mathbf{Z}_{2}$ symmetry ensures the absence of flavor changing neutral currents (FCNCs) at tree level, there is a potential worry about loop induced effects.  However, because of the minimal flavor violating coupling, we can avoid large FCNCs if  the mass terms that break $\mathbf{S}_2$ are approximately diagonal in flavor space.  For the up sector, the relevant soft mass terms are:
\be
\mathcal{L}_{\rm soft} = m_Q Q Q^c + m_U U U^c.
\ee
These masses softly break the exchange symmetry between $h$ and $\phi$, and a splitting will be induced at the loop level.  It is dominated by the top Yukawa coupling, and is finite at one-loop.  This is because both the $\mathbf{S}_2^A$ and $\mathbf{S}_2^B$ must be broken in order for $h$ and $\phi$ to be split.\footnote{When there are only doublet or singlet partners, the mass splitting is logarithmically sensitive to the cut-off.}
\be
\label{eqn:splitting}
m_\phi^2 - m_h^2 =  + \frac{3 \lambda_{\rm top}^2}{8 \pi^2} \frac{m_Q^2 m_U^2}{m_Q^2 - m_U^2} \log \left(\frac{m_Q^2}{m_U^2} \right).
\ee
This is a technically natural mass splitting.  At higher-loop order, the mass splitting is at worst logarithmically sensitive because the spurions $m_Q$ and $m_U$ are dimensionful.   Because $h$ and $\phi$ have the same Standard Model quantum numbers, gauge interactions do not split $h$ and $\phi$ until electroweak symmetry is broken.  The sign of $(m_\phi^2 - m_h^2)$ is positive, which allows for the Higgs (mass)$^2$ to be negative but the $\phi$ (mass)$^2$ to be positive.

Of course, the above discussion assumes that whatever dynamics generates the masses for the partner fermions has the same primordial $\mathbf{S}_{2}$ symmetry as the Higgs potential.   The breaking should only appear after an $\mathbf{S}_{2}$-breaking condensate forms.  However, the dynamics of the $\mathbf{S}_{2}$-breaking need not be visible near the electroweak scale.  For example, if $Q$ and $Q^c$ couple to technifermions $\psi$ and $\psi^c$ through a dimension six operator $\psi\psi^c Q Q^c / M_{\rm Pl}^2$, then the condensation scale $\langle \psi \psi^c \rangle \equiv \Lambda_{\rm TC}^3 \sim m_Q M_{\rm Pl}^2$.

\begin{figure}
\begin{center}
\includegraphics[width=\columnwidth]{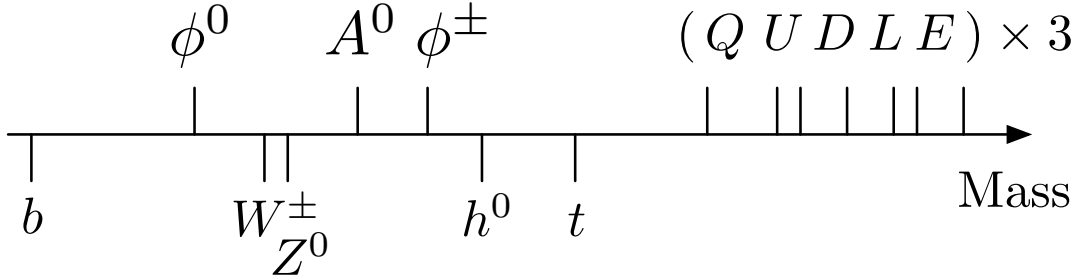}
\end{center}
\caption{\label{fig:spectrum}A typical spectrum for a technically natural scalar Dark Matter model.  The new $\phi$ doublet resolves into three distinct states, a neutral scalar $\phi^{0}$, a pseudo-scalar $A^{0}$, and a charged scalar $\phi^{\pm}$.  These are the lowest lying beyond the Standard Model states.  The partners to the Standard Model fermions are typically a loop factor heavier, though numerically this splitting may not be too large.  The partners of the top quark are expected to be at several hundred GeV.}
\end{figure}

After electroweak symmetry breaking, the quartic couplings in \eq{eqn:scalarpotential} lead to splittings among the components of $\phi$.  The scalar Dark Matter particle $\phi^0$ is accompanied by a pseudo-scalar field $A^0$ and a charged field $\phi^\pm$ with masses
\beq{eq:masssplit}
m_{A^0}^2 = m_{\phi^0}^2 - \lambda_5 v^2, \qquad m_{\phi^\pm}^2 = m_{\phi^0}^2 - \frac{v^2}{2}(\lambda_4 + \lambda_5), 
\ee  
where $v \approx 246 \mbox{ GeV}$ is the Higgs vacuum expectation value (vev).  Note $m_{\phi^0}^2$ gets contributions from the fine-tuned bare mass, the Higgs/Dark Matter splitting in \eq{eqn:splitting}, and the $\lambda_i$ quartics.  A typical spectrum is given in Fig.~\ref{fig:spectrum}.  We will take the real scalar to be the lightest field, which will require $\lambda_{5} <0$.  As we will see, the Dark Matter relic abundance depends on the coupling of $\phi^0$ to the physical Higgs $h^0$:
\beq{eq:lambdaDM}
\mathcal{L}_{\rm DM} = - \frac{\lambda_{\rm DM} v }{2} \left( \phi^0 \right)^2 h^0, \qquad \lambda_{\rm DM} = \lambda_3 + \lambda_4 + \lambda_5.
\ee
To ensure the stability of the vacuum, 
\begin{equation}
\label{eqn:vaccumstability}
\lambda_1 >0, \qquad \lambda_{3} > -2 \lambda_{1}, \qquad \lambda_{\rm DM}=\lambda_{3} +\lambda_{4} - |\lambda_{5}| > -2 \lambda_{1}.
\end{equation}
We will concentrate on the region $\lambda_{3}, \lambda_{\rm DM} > 0$ where these conditions are trivially satisfied.

\section{Dark Matter Properties}
\label{sec:dm}
\subsection{Thermal Relic Abundance}
\label{sec:relic}
As long as the partner fermions are sufficiently heavy, the thermal relic abundance calculation is essentially identical to the one recently presented in \cite{Inert} and qualitatively discussed in \cite{Lawrence}.  There are two qualitatively different regions.  First, for $\phi^0$ masses greater than the $W$ mass, there is efficient annihilation into $W$ pairs via the t-channel exchange of the $\phi^{\pm}$.  In this case, agreement with the observed relic density from cosmological measurements \cite{PDG} forces the $\phi^0$ mass into the multi-TeV region (see \cite{StrumiaDM} for discussion of doublet Dark Matter in this case).  \eq{eqn:splitting} would then indicate that the heavy colored partners would have masses a loop factor larger, making observation in the near term impossible.  Thus, we will concentrate on the second region, $m_\phi <m_W$.  Even in this regime, there is the possibility of a too-efficient co-annihilation of $\phi^0$ with $A^0$ via a $Z$ boson, or the co-annihilation of $\phi^0$ with $\phi^{\pm}$ via a $W$ boson.  Absent these co-annihilation processes, there is be a perfectly healthy region with reasonable Dark Matter abundance.

\begin{figure}
\begin{center}
\includegraphics[width=10cm]{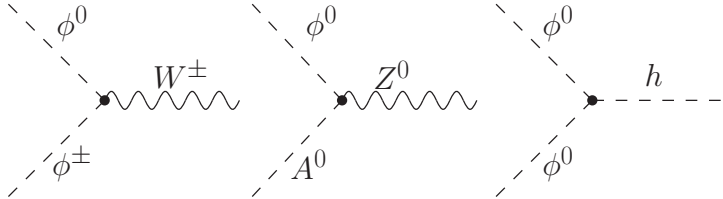}
\caption{\label{Fig:annihilation}
Diagrams that potentially contribute to the annihilation of the Dark Matter candidate $\phi^{0}$ in the regime $m_{\phi} < m_{W}, m_{h}$.  In the text, we assume that the splittings $\Delta m _{\phi^{0} A}$ and $\Delta m_{\phi^{0} \phi^{\pm}}$ are large enough to suppress the first two co-annihilation diagrams shown here. The third diagram dominates the relic abundance calculation. There are also diagrams with heavy fermions on the $t$-channel, but these are typically negligible.}
\end{center}
\end{figure}

Fortunately, the symmetries of this model allow precisely the set of operators that are expected to suppress this dangerous co-annihilation.  As shown in \eq{eq:masssplit}, after the Higgs boson gets a vev, the last two terms terms in \eq{eqn:scalarpotential} split the $\phi^{0}$, $A^{0}$ and $\phi^{\pm}$ masses.  The amount of splitting is set by the electroweak vev, and can easily be tens of GeV for modest values of the $\lambda_{i}$.  The co-annihilation effects will be suppressed by factors of $e^{-\Delta m_{\phi}/T_{f}}$, with the freeze-out temperature $T_{f} \sim m_{\phi} /20$.  This is sufficient to render them negligible.

\begin{figure}[p]
\begin{center}
\begin{tabular}{c}
\hspace{-.5cm} \includegraphics[width=13cm]{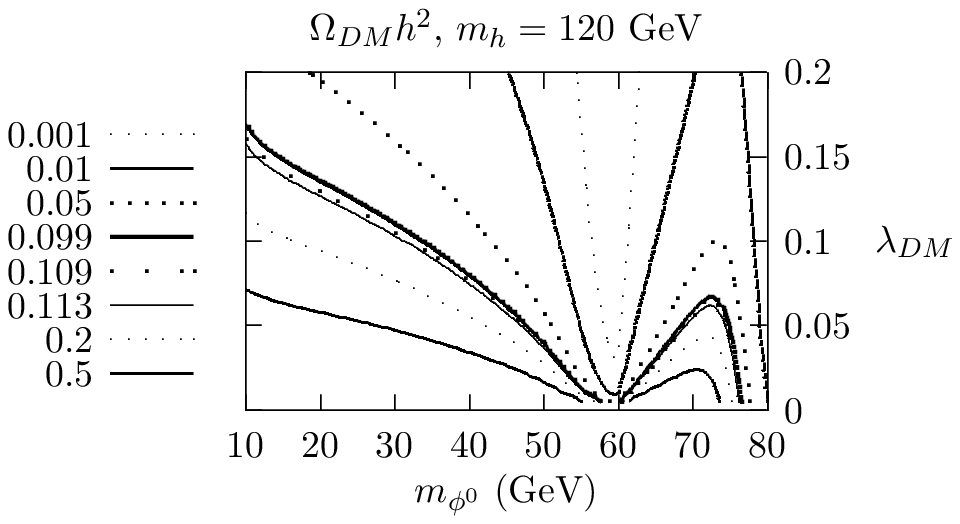} \vspace{-.5cm} \\
\\
\hspace{-.5cm} \includegraphics[width=13cm]{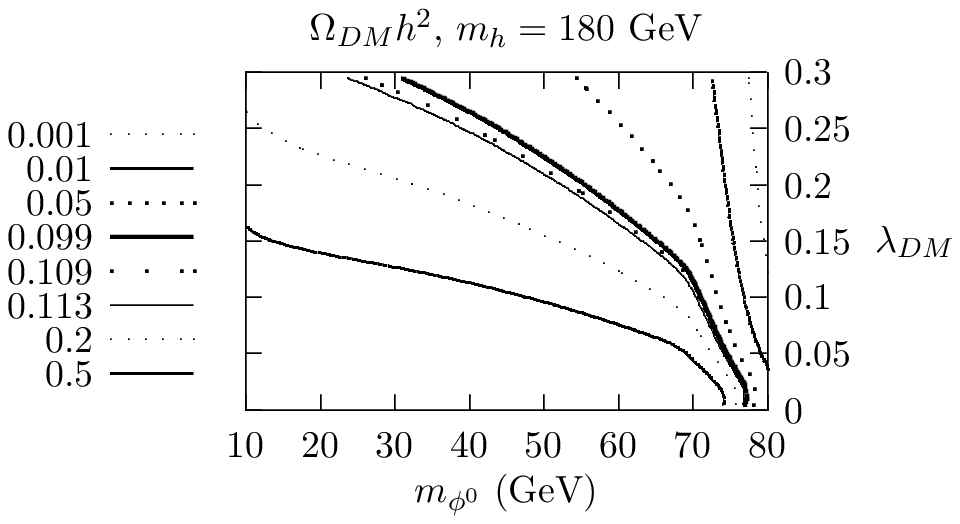}
\end{tabular}
\caption{\label{Fig:relic}
Relic abundance contours of $\Omega_{DM} h^{2}$ for the the $\phi^0$ field as a function of the Dark Matter mass $m_{\phi^0}$ and its coupling to the Higgs boson $\lambda_{\rm DM}$.  Current experiments bound the total Dark Matter abundance $0.099 < \Omega_{DM} h^{2} < 0.113$ \cite{PDG}.}  
\end{center}
\end{figure}

In the following, we assume that the $\lambda_{i}$ are sufficiently large to completely negate the co-annihilation effects.  This is also largely favored by the direct collider limits from the Large Electron Positron collider (LEP) (see Sec.~\ref{sec:directcollider}).  To calculate the relic abundance we use the \texttt{micrOMEGAs} \cite{MM} program.  In version 2.0, it is straightforward to implement a new Lagrangian at tree level using the \texttt{LanHEP} \cite{LanHEP} utility.  This approach is also the one taken by \cite{Inert} for the Inert Doublet model, and our results appear consistent with theirs. The dominant annihilation is via an s-channel Higgs boson (see Fig.~\ref{Fig:annihilation}).  

In Fig.~\ref{Fig:relic} we show the relic abundance contours for two different choices of the Higgs boson mass, $m_h =120$ GeV and $m_{h} = 180$ GeV.  We have plotted the contours as a function of the Dark Matter mass $m_{\phi^{0}}$ and the coupling of the Higgs boson to the Dark Matter $\lambda_{DM}$ as in \eq{eq:lambdaDM}.
Changing the Higgs boson mass has two clear effects.  First, a heavier Higgs boson generically suppresses the annihilation cross section, and thus calls for larger values of $\lambda_{DM}$ to replicate the observed relic abundance, $0.099 < \Omega_{DM} h^{2} < 0.113$ \cite{PDG}. As shown in the figure, for a lighter Higgs boson mass, one wants a somewhat small $\lambda_{DM} \lsim 0.1$.  For a heavier Higgs mass, larger values can be accommodated. Second, the Dark Matter relic abundance drops dramatically when $m_{\phi^{0}} \approx m_{h}/2$ due to resonant annihilation.  While this pole is clearly visible in the plot for $m_{h} = 120$ GeV, for a Higgs boson mass of 180 GeV this pole is in the region where efficient annihilation into $W$ boson pairs is already available. 

\subsection{Dark Matter Direct Detection}
\label{sec:dd}
Direct detection experiments force a viable $\phi^{0}$ Dark Matter candidate to have a small splitting from the $A^{0}$.  This goes beyond the co-annihilation considerations discussed above, and even applies in the multi-TeV $\phi^{0}$ case.  The $\phi^{0}$--$A^{0}$--$Z$ coupling can lead to a large coherent scattering off of nuclei, which is by now firmly excluded in the range that gives the correct relic abundance by, for example, the Cryogenic Dark Matter Search (CDMS) experiment \cite{CDMS}.  In fact, in the absence of a $\phi^{0}$--$A^{0}$ splitting, the Dark Matter behaves identically to pure left-handed sneutrino Dark Matter, which is well known to be excluded on similar grounds.  A splitting that exceeds the momentum transfer (q $\approx$ 10 keV) in a $\phi^{0}$--nucleon collision is sufficient to shut off this coupling in direct detection experiments.  In the low mass window, we required a splitting of order 10 GeV to shut off co-annihilations, so the $\phi^{0}$--$A^{0}$--$Z$ coupling can be safely neglected in direct detection experiments.

In fact, in the window of interest, the direct detection cross section is completely dominated by the exchange of the Higgs boson.  The formula for the spin-independent scattering off a nucleus ${\mathcal N}$ becomes particularly simple and has been presented for an equivalent model in Ref.~\cite{Lawrence}:
\begin{equation}\sigma = \frac{m_{r}^{2}}{4 \pi} \left(\frac{\lambda_{DM}}{m_{\phi^{0}} m_{h}^{2}}\right)^{2} f^{2} m^2_{\mathcal N}, 
\end{equation}
where $m_{r}$ is the reduced mass of the $\phi$ and nucleon, and $f \sim$ 0.3 is the matrix element
\be
\langle {\mathcal N} | \sum{m_{q}} q \bar{q}| {\mathcal N} \rangle= f m_{\mathcal N} \langle {\mathcal N}|{\mathcal N} \rangle.
\ee
A typical point in the region allowed by the relic abundance---$m_{\phi}=40 \mbox{ GeV}$, $m_{h}=120 \mbox{ GeV}$, and $\lambda_{DM}=0.1$---has a cross section for scattering off a proton $\sigma_{\phi p}=6 \times 10^{-8}$ pb.  This is roughly two orders of magnitude below the current limit from CDMS \cite{CDMS}.  The next generation of experiments may reach this level of sensitivity.  Of course, the uncertainties from the matrix elements and the local Dark Matter density make the accessibility of this cross section somewhat uncertain.

\section{Collider Considerations}
\label{sec:collider}

\subsection{Direct Collider Bounds}
\label{sec:directcollider}
While there is no existing experimental search that can be directly translated into a search for the $\phi$ fields, the search for charginos and neutralinos is very close.  For example, pair production of $\phi^{\pm}$ with a decay via (likely off-shell) $W$ bosons to the $\phi^{0}$ presents a identical topology to the production of chargino pairs that decay to a stable neutralino.  While the LEP bounds on a chargino are near the kinematic limit 103.5 GeV \cite{LEPChargino}, the bound on the $\phi^{\pm}$ will be weaker, simply because the $\phi^{\pm}$ has a smaller production cross section than charginos.  

This decreased cross section is a result of two factors. First, there is the standard factor of four difference when translating from a fermion to a scalar.  Second, the phase space suppression for scalar production persists longer than for fermions.  The combination of these two factors can represent a decrease in the production cross section of an order of magnitude near masses of 100 GeV.  Depending on the mass of the neutralino, the OPAL collaboration places cross section bounds between between 0.1 and 0.3 pb \cite{Opal1,Opal2}.  Converting this to a mass bound on $\phi^\pm$:
\begin{equation}
m_{\phi^{\pm}} \gsim 70 \mbox{ -- } 90 \mbox{ GeV}, \qquad  e^{+} e^{-} \rightarrow \phi^{+} \phi^{-}.
\end{equation}
Since this limit is luminosity limited, it could likely be improved by combining the data sets from the various LEP experiments.\footnote{Note that these limits are weakened in the limit where $m_{\phi^{\pm}}$ is very close to $m_{\phi^{0}}$, but we do not expect to obtain this spectrum.  This would lead to rapid co-annihilation through the $W$, and consequently a too-low relic abundance for the Dark Matter.}  In addition, one might consider the limits from, e.g., slepton searches.  The sleptons have nearly identical production cross section as the $\phi$ fields.  However, the branching ratio of the $\phi$ field to leptons follows that of the off-shell $W$, so the limits derived from these searches are substantially weaker than those found from the chargino searches.

There is also a bound that comes from the process $e^{+} e^{-} \rightarrow A^{0} \phi^{0}$ at LEP.  Again, direct translation is difficult, but comparison to the neutralino associated production $e^{+} e^{-} \rightarrow \chi_2^{0} \chi_1^{0}$ provides a good guide.  The L3 collaboration has presented their limits \cite{L3} as a function of mass and cross section.  Over much of the relevant range the bound on the production cross section is $\sigma_{\phi A} < 0.2$ pb.  If the mass splitting between the $\phi^0$ and $A^0$ is sufficiently small, this bound can be significantly degraded.  However, the requirement that we do not want significant co-annihilation between $\phi^0$ and $A^0$ largely eliminates this consideration. As a rough rule of thumb, this bound on the cross section forces the summed mass of the $\phi$ and $A$ to be greater than 130 GeV.  

\begin{figure}
\begin{center}
\includegraphics[width=9cm,angle=-90]{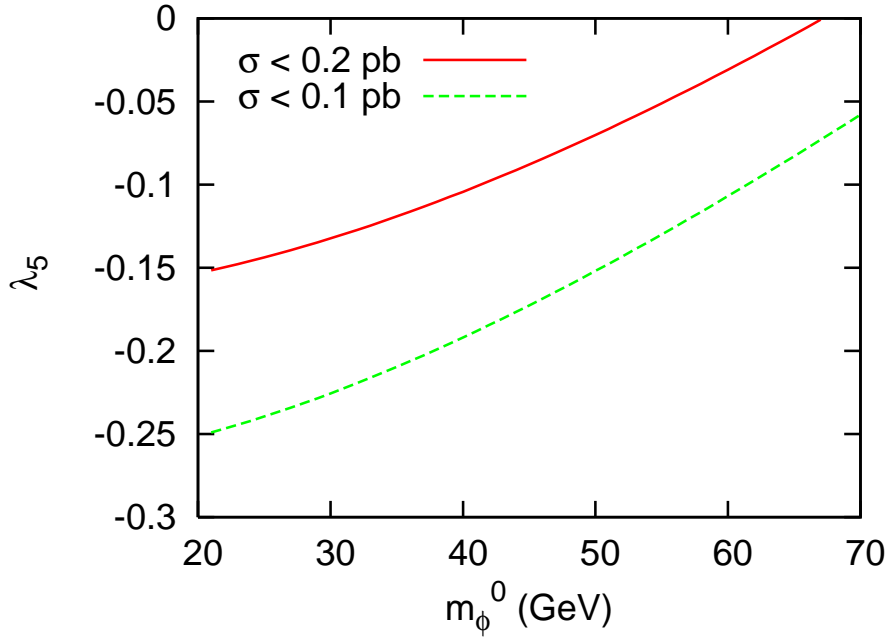}
\caption{\label{fig:l5}
The maximum value of $\lambda_{5}$ assuming two different bounds on the production cross section $\sigma_{\phi^{0} A^{0}}$ from LEP.  $\lambda_{5}$ must be more negative than each of the given curves to achieve a sufficient splitting between $\phi^{0}$ and $A^{0}$. A comparison with the necessary $\lambda_{DM} \equiv \lambda_{3} +\lambda_{4} + \lambda_{5}$, in Fig.~3 gives an indication of the amount of cancellation necessary between the couplings.  Unless $m_{\phi^{0}}$ is quite small, it is minimal.}
\end{center}
\end{figure}

If $m_{\phi}$ is quite light, say $m_{\phi} \approx 30$ GeV,  the requisite splitting between $\phi^0$ and $A^0$ can only be achieved if $\lambda_{5}$ is sizable (and negative).  Since the coupling of Dark Matter to the Higgs is controlled by the combination $\lambda_{\rm{DM}}$ of \eq{eq:lambdaDM}, achieving the correct relic abundance in this case will require some cancellation between the various $\lambda_{i}$.  This tension is ameliorated if $m_{\phi}$ is somewhat larger (but still less than $M_{W}$).  To get a feel for the size of the $\lambda_{5}$ required by the collider bounds, we have plotted the necessary  $\lambda_{5}$ as a function of $m_{\phi}$, assuming two different bounds on the production from LEP, see Fig.~\ref{fig:l5}.   

Finally, it is no surprise that the heavy fermions have not yet been observed.  As we will discuss in detail below in \eq{eqn:Qmass}, the expected mass scale for these particles is $M \approx 500 \mbox{ GeV}$, too heavy for observation at the Tevatron. We discuss their production at the LHC in Sec.~\ref{sec:lhc}.

\subsection{Precision Electroweak Bounds}

Whenever there are new particles that get some of their mass from electroweak symmetry breaking, there is a possibility for modifications of the precision electroweak parameters.  The $\mathbf{S}_2$ symmetries in Eqs.~(\ref{eqn:xchange1}) and (\ref{eqn:xchange2}) force the existence of the $\lambda_{\rm top} Q h U^c$ coupling in \eq{eqn:yukawa}, which gives a violation of custodial $SU(2)$.  Therefore, there are contributions to the $T$ parameter in this model, which will constrain how light $m_Q$ and $m_U$ can be.  As we will see, considerations of the $S$ parameter and $Z \rightarrow b \bar{b}$ generically do not impose additional constraints.  In the regions of interest, the contributions of $\phi$ to precision electroweak constraints are sub-dominant. 

In the limit that either $m_Q$ or $m_U$ is very large, the mass of the new particles is dominantly vector-like, and the modification to $T$ will vanish.  That is, these modifications to $T$ decouple.  Maximal modifications occur when $m_Q = m_U$, and we will assume this relation for deriving bounds.  We denote the common vector-like mass by $M$.  The mass matrix in the top sector is
\beq{eq:Tmassmatrix}
\left(\begin{array}{cc}Q_t & T \end{array}\right)
\left(\begin{array}{cc}M & m_{\rm top} \\0 & M\end{array}\right)
\left(\begin{array}{c}Q^c_t \\ T^c \end{array}\right),
\ee
where $Q_t$ is the top component of the $Q$ doublet.  This mass matrix is similar to the mass matrix in the top sector of Universal Extra Dimension models \cite{UED}, and the precision electroweak constraints share qualitative features with these models.  

Expanding in $m_{\rm top}/M$, the mass eigenvalues are
\beq{eq:Tmasssplit}
m_\pm = M \left( 1 \pm \frac{1}{2}\frac{m_{\rm top}}{M} + \frac{1}{8} \frac{m_{\rm top}^{2}}{M^{2}} - \frac{1}{128} \frac{m_{\rm top}^{4}}{M^{4}}+ \cdots \right).
\ee
The mass eigenstates $T_\pm$, $T_\pm^c$ are
\be
\left(\begin{array}{c}T_+ \\T_-\end{array}\right) = \left(\begin{array}{cc}\cos \theta & \sin \theta \\ -\sin \theta & \cos \theta \end{array}\right)\left(\begin{array}{c}Q_t \\ T\end{array}\right), \quad \left(\begin{array}{c}T^c_+ \\T^c_-\end{array}\right) = \left(\begin{array}{cc}\cos \theta^c & \sin \theta^c \\ -\sin \theta^c & \cos \theta^c \end{array}\right)\left(\begin{array}{c}Q^c_t \\ T^c\end{array}\right),
\ee
where
\be
\cos \theta = \sin \theta^c = \frac{1}{\sqrt{2}} + \frac{1}{4\sqrt{2}} \frac{m_{\rm top}}{M} - \frac{1}{32\sqrt{2}} \frac{m_{\rm top}^2}{M^2} - \frac{3}{128\sqrt{2}} \frac{m_{\rm top}^3}{M^3} 
 + \cdots,
\ee
\be
\cos \theta^c = \sin \theta = \frac{1}{\sqrt{2}} - \frac{1}{4\sqrt{2}} \frac{m_{\rm top}}{M} - \frac{1}{32\sqrt{2}} \frac{m_{\rm top}^2}{M^2} + \frac{3}{128\sqrt{2}} \frac{m_{\rm top}^3}{M^3} 
 + \cdots.
\ee
These mixing angles can be used to calculate the couplings of $T_\pm$ and $T^c_\pm$ to Standard Model gauge bosons.  To leading order in $m_{\rm bottom} / m_{\rm top}$, the mass of the the bottom component $Q_b$ of the $Q$ doublet is just $m_Q$ and the mixing with $B$ can be ignored.  

With this information, we can use the standard techniques of \cite{PeskinTakeuchi} to calculate the $T$ parameter.  
\be
\alpha T \simeq \frac{7 G_F}{80 \sqrt{2} \pi^2}\frac{m_{\rm top}^4}{M^2}\left(1  - \frac{13}{49} \frac{m_{\rm top}^2}{M^2} + \cdots  \right)
\ee
We assume that $m_{H} \approx 120$ GeV, and as is conventional fix $U \simeq 0$. In this case, $T= -0.03 \pm 0.09  $ \cite{PDG}, and even at the $68\%$ confidence level, this puts the very mild limit of
\beq{eqn:Qconstraint}
M \gsim 360 \mbox{ GeV}.
\ee
Of course, a small positive contribution to $T$ can be compensated by a heavier Higgs boson.  In this case, the bounds from $T$ can be relaxed, and the bound from the $S$ parameter may be the strongest.

We calculate the $S$ parameter using the methods of \cite{PeskinTakeuchi}.  The result is:
\be
S \simeq \frac{2}{5\pi} \frac{m_{\rm top}^2}{M^2} \left(1 - \frac{11}{28} \frac{m_{\rm top}^2}{M^2} + \cdots  \right),
\ee
Once \eq{eqn:Qconstraint} is satisfied, the $S$ parameter is smaller than the experimental limit $S=-0.07 \pm 0.09$ \cite{PDG}, where we have again fixed $U=0$.  

The only potentially significant non-oblique correction is to the coupling of the $Z$ boson to the left-handed $b$ quark.  Loops of heavy top quarks and charged $\phi$ fields can give a modification that is proportional to the top Yukawa coupling.  Again, the effect decouples as the vector-like masses become large. The coupling of the $Z$ to the left-handed $b$ quark is modified as:
\be
{\mathcal L_{\rm eff}} = \frac{e}{\sin{\theta_W} \cos{\theta_W}} Z_{\mu} \bar{b} \gamma^{\mu} (g_{L} + \delta g_{L}) P_{L} b
\ee
where $P_{L}$ is the left-handed projection operator, $(1-\gamma_{5})/2$, and $\theta_{W}$ is the weak mixing angle.  The corrections to $g_L = -1/2$ are  
\be
\delta g_L \simeq 
-\frac{m_{\rm top}^2}{64 \pi^2 v^2} \frac{m_{\rm top}^{2}}{M^{2}}
 \left( 1 - \frac{4}{15} \frac{m_{\rm top}^{2}}{M^{2}} + \frac{2}{3} \frac{m_{\phi^{\pm}}^{2}}{M^{2}} + \cdots \right)
\ee
with $v \simeq 246$ GeV.  With the above definition, $R_{b} \approx R_{b}^{SM}( 1-3.56 \, \delta g_{L})$ \cite{BurgessRb}.  This quantity has been measured to the $7 \times 10^{-4}$ level \cite{PDG}.  So, once the constraint of \eq{eqn:Qconstraint} is imposed, the correction is safely within experimental errors.

\subsection{Collider Signatures at the LHC and Beyond}
\label{sec:lhc}

The phenomenology of this model is quite similar to the top sector of little Higgs with $T$-parity \cite{Cheng:2005as,Belyaev:2006jh}, or even that of Universal Extra Dimensions \cite{Cheng:2002ab}.  There are heavy fermions that decay to a stable $\mathbf{Z}_{2}$-odd particle that escapes as missing energy.  However, unlike any of the models that attempt to the solve the hierarchy problem, there are no partners of the gauge bosons.  In particular, there is no analog of Kaluza-Klein gluons or gluinos, so observations of color octet new particles would exclude the model considered here.  While there are no ``electroweakinos'' either, the $\phi$ field behaves in a manner quite similar to a chargino and a pair of neutralinos (\emph{i.e.} a Higgsino), so disentangling this sector may be challenging.

The Higgs boson phenomenology of this model is similar to that of the Inert Doublet Model.  In particular, there is the possibility of Higgs decays to the various components of the $\phi$ multiplet.  Unless there is a contribution to the $T$ parameter to allow for a heavier Higgs, the LEP bounds likely indicate that only $h \rightarrow \phi^{0} \phi^{0}$ decays are possible.  In this case, the Higgs undergoes invisible decays with \cite{Lawrence}
\be
\Gamma_{\rm invisible} = \frac{v^{2}}{32 \pi m_{h}} \lambda_{DM}^{2} \left(1- \frac{4m_{\phi}^{2}}{m_{h}^{2}}\right)^{1/2}.
\ee
For the model at hand, we have insight into the size of $\lambda_{DM}$ as a function of $m_{\phi}$ from the relic abundance calculation of Sec.~\ref{sec:relic}.  If the Higgs is below the $WW$ threshold, this can dominate the Higgs boson width, so the Higgs would be discovered via the invisible mode.  In the case where the Higgs is above the $W$ threshold, the decays to $W$ pairs will likely dominate, but a non-negligible fraction ($\approx$ 20\%) of decays may still be invisible.  It will be a challenge to observe this mode at the LHC, but searches using the Weak Boson Fusion production mode are promising \cite{WBFInvis}.  Detailed detector studies show that invisible decays of the Higgs boson may be excluded at the 95\% confidence level down to branching ratios of $\approx 15 \%$ at CMS and $\approx 35\%$ at ATLAS \cite{CMSATLAS}.

\begin{figure}
\begin{center}
\includegraphics[angle=-90,width=9.5cm]{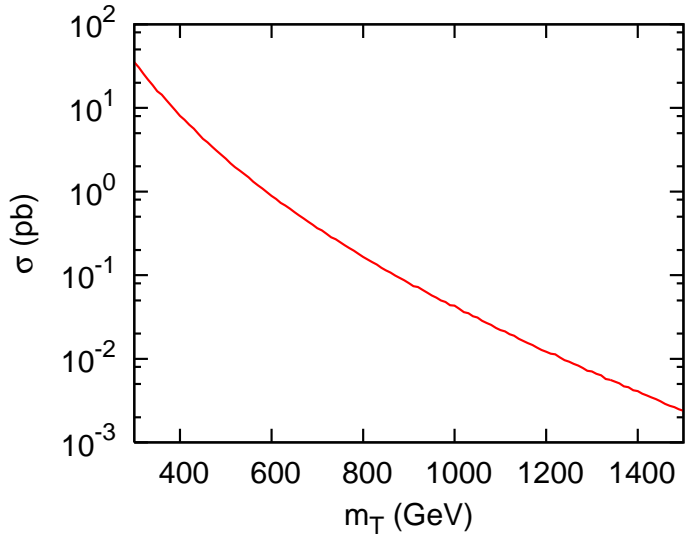}
\caption{\label{Fig:txsec}
Production cross section for the top partner at the LHC, $\sqrt{s} = 14$ TeV.}
\end{center}
\end{figure}

Since the mass of the top partner feeds directly into the mass of the $\phi$ field, the prediction for its mass is rather firm.  Using the result of \eq{eqn:splitting}, and setting $m_{Q} \approx m_{T}$ we find:
\begin{equation}
\label{eqn:Qmass}
m_{T} = \sqrt{ \frac{8 \pi}{3 \lambda_{\rm top}^{2}} 
 \left( m_{\phi^{0}}^{2} - \frac{\lambda_{3} v^{2}}{2} + \frac{m_{h}^{2}}{2} \right) }
\end{equation}
For typical values, $m_{\phi}=70$ GeV, $m_h= 120$ GeV, and $\lambda_{3}=.1$, we find $m_{T} = 650$ GeV.  This naive expectation for the mass is consistent with the derived precision electroweak bounds from \eq{eqn:Qconstraint}.  In most cases, the sum on the right-hand side is dominated by the term that depends on the Higgs boson mass.  In this case, we have
\begin{equation}
\label{eqn:QmassSimp}
m_{T} \approx 620 \mbox{ GeV} \left(\frac{m_{h}}{120 \mbox{ GeV}}\right).
\end{equation}
We plot the cross section for $T$ quark production at the LHC in Fig.~\ref{Fig:txsec}.  The cross section was calculated using {\tt Pythia 6.4} \cite{Pythia}, with the CTEQ5L Parton Distribution.  The calculation is identical to a vector-like generation of heavy quarks.
 
There are a few different decay modes possible for the $T$ quark.  Minimally $T \rightarrow t \phi^0$, which yields the same production and decay topology as the general models considered in \cite{Meade:2006dw}.  Because the $\phi$ comes in a complete $SU(2)_L$ multiplet, there will also be cascade decays like $T \rightarrow b \phi^+ \rightarrow b W^+ \phi^0$.  As we saw in \eq{eq:Tmassmatrix}, the Higgs vev introduces mixing between the doublet and singlet $T$ partners.  If mixing angle is large enough and there is available phase space, then there is the possibility of decays like $T_+ \rightarrow Z^0 T_-$ and $T_+ \rightarrow h T_-$.  

A crucial component of verifying this model is measuring the Yukawa coupling of the heavy top partners to the $\phi$ field.  It is difficult to imagine how one might do this at the LHC.  At a future linear collider, a threshold scan might allow a determination of the heavy top quark's width if it were kinematically accessible.  It is amusing that this Yukawa coupling is equal to the top Yukawa, even though there is no attempt to solve the hierarchy problem in this model.

\section{Other Symmetry Possibilities}
\label{sec:othersym}

In Sec.~\ref{sec:themodel}, we presented a model that realizes scalar Dark Matter without an additional fine-tuning by relating the Dark Matter to the Higgs with a discrete symmetry.  We now explore alternative symmetry structures.

There are two separate approaches for relating the Higgs to Dark Matter.  One possibility, quite distinct from Sec.~\ref{sec:themodel}, is to attempt to construct a theory where the $h$ and the $\phi$ have different quantum numbers.  This path utilizes an extended symmetry group $G_{\rm ext}$ that completely contains the Standard Model gauge group $G_{\rm SM} = SU(3)_C \times SU(2)_L \times U(1)_Y$.   The Dark Matter would arise from the extra components of the Higgs multiplet that now fill out a complete representation of $G_{\rm ext}$.    This kind of construction is familiar from composite Higgs theories.  

However, models with $G_{\rm SM} \supset G_{\rm ext}$ are unlikely to be phenomenologically viable, once we impose the constraint of not introducing additional fine-tunings.  This requirement suggests that $G_{\rm ext}$ should be broken to $G_{\rm SM}$ by some form of dimensional transmutation, e.g. a technicolor-like theory.  The Dark Matter particle will receive a splitting from the Higgs boson scale of order $(\mbox{loop factor}) f_{\rm{ext}}$, where $f_{\rm ext}$ is the scale of the $G_{\rm ext}$ gauge-breaking.  Only above this scale are the gauge interactions of the Higgs and Dark Matter identical.  To make the splitting small enough, this scale should not be much more than a TeV.   Meanwhile, fermions will also come in complete representations of $G_{\rm ext}$.  Heavy masses for the partner fermions will also induce splittings between the Higgs and the Dark Matter, just as in Sec.~\ref{sec:themodel}.  However, these fermion masses are usually suppressed  by factors of $f_{\rm ext}/f_{\rm ETC}$, where $f_{\rm ETC}$ parametrizes an ``extended-technicolor'' scale.  So, once $f_{\rm ext}$ is taken small enough to avoid a too large correction to the Dark Matter (mass)$^{2}$, there is a real possibility that the partner fermions are too light and would have already been observed.  So, while it may not be impossible to build a model of this type, the considerations mentioned here seem to disfavor this possibility.

Therefore, if we want technically natural scalar Dark Matter with new colored particles at the TeV scale, we are pushed towards theories where $h$ and $\phi$ have the same quantum numbers. This is most easily implemented with a product group structure $G_{\rm ext} = G_{\rm SM} \times G_{\rm new}$.  Regardless of whether $G_{\rm new}$ is discrete or continuous, gauged or global, only the breaking of $G_{\rm new}$ in the fermion sector is relevant for the Higgs/Dark Matter splitting, as the Higgs and Dark Matter would have the same gauge quantum numbers.    

Below, we show how one can use a continuous symmetry that commutes with the Standard Model to relate the Higgs to Dark Matter.  Then, we attempt to use the well known observation that the Standard Model with six Higgs doublets approximately unifies at $10^{15}$ GeV \cite{Willenbrock} to attempt create a technically natural scalar Dark Matter model with gauge coupling unification, in the same spirit as Ref.~\cite{Friendly,Senatore}.  Existing experimental constraints seem to disfavor achieving gauge coupling in this way, though one can use discrete symmetries in four- or eight-doublet models to get rough gauge coupling unification.

\subsection{New Continuous Symmetries?}

If only models with the symmetry structure $G_{\rm SM} \times G_{\rm new}$ are likely to be viable, what are the possibilities for $G_{\rm new}$?  Continuous symmetries are much more constraining than discrete symmetries, and so will be much more constrained phenomenologically.  As an example, consider the symmetry $G_{\rm new}= SO(2)$.  We can place the Standard Model Higgs $h$ and the Dark Matter particle $\phi$ in a single multiplet $\Phi= (h, \phi)$.  The most general form of the potential in this case is 
\begin{equation}
\label{eqn:altscalarpotential}
V(\Phi)= m^{2} |\Phi|^{2} + \kappa_{1} |\Phi|^{4} + \kappa_2 | \epsilon^{ij} \epsilon_{ab} \Phi_i^a \Phi_j^b|^2 + \kappa_{3} (\epsilon^{ij} \Phi^{\dagger}_{i} \Phi_{j})^2 ,
\end{equation}
where $a,b$ are $SU(2)_{L}$ indices, and $i,j$ are the indices of the $SO(2)$ $\Phi$ doublet.  In this model, it is obvious that the Dark Matter is light because it is a pseudo-Goldstone boson of spontaneous $SO(2)$ breaking.  The Dark Matter then gets a mass from the soft breaking of the $SO(2)$ symmetry from the heavy quark masses.

After electroweak symmetry breaking, $\kappa_{2,3}$ give terms that split the various $\phi$ fields.  Going back to the notation of \eq{eqn:scalarpotential}, we have
\be
\lambda_1 = \kappa_1 , \quad \lambda_3 = 2 \kappa_1 + 4 \kappa_2, \quad \lambda_4 = -4 \kappa_2 - 2 \kappa_3, \quad \lambda_5 = 2 \kappa_3.
\ee
In this scenario, the physical Higgs and $\phi$ masses are related to the $h$--$\phi$--$\phi$ couplings, so the masses and the relic abundances are no longer independent.  From the above expression, we find $\lambda_{DM} = 2 \kappa_{1}$.  This coupling is related to the Higgs boson mass, $\kappa_{1} = \frac{1}{2} m_{h}^{2}/ v^{2}$.  Thus, for a 120 GeV Higgs boson, the $SO(2)$ symmetry forces $\lambda_{DM} \approx .25$, and for a 180 GeV Higgs boson $\lambda_{DM} \approx .54$.  Examining the plots of Sec. \ref{sec:relic}, it is clear that this large of a coupling will lead to too little Dark Matter for these cases.  Because of the relationship between $\kappa_{1}$ and $m_h$, moving to larger Higgs masses will not help to increase the relic abundance.

Other continuous symmetry possibilities are even more constrained.  In order for the Dark Matter to be light, we had to split the components of $\phi$ to avoid aggressive co-annihilation and direct direction, and in particular, $\lambda_5$ had to be non-zero.  If $G_{\rm new}= SU(2)$, then the $\kappa_3$ term in \eq{eqn:altscalarpotential} would be forbidden, shutting off the splitting between the real and imaginary neutral $\phi$ scalars.  Without the $\lambda_5$ term, direct detection limits exclude scalar doublet Dark Matter with the mass favored by the thermal relic abundance calculation.  So, while it may be possible to build more baroque models utilizing continuous symmetries, the two simple examples presented here help to illustrate the difficulties encountered when attempting to do so.     

\subsection{Gauge Coupling Unification?}
\label{sec:unification}
Is it possible to get gauge coupling unification with natural scalar Dark Matter?  If so, the analogy between this model and Split SUSY would be strengthened.  As long as there are the same number of doublet and singlet partners of the Standard Model fermions, then the fermion sector will come in complete $SU(5)$ multiplets.  Then, if the discrete symmetry group $G$ yields six light Higgs-like doublets---some of which are $\mathbf{Z}_2$-odd to be stable Dark Matter---then the Standard Model gauge couplings will unify within typical GUT scale thresholds \cite{Willenbrock}.  While it is possible to write down reasonable Lagrangians that have this property, we will see that they generically have problems with tree-level flavor changing neutral currents.  Other possibilities where the fermions do not come in a complete GUT multiplet may exist, but seem to require the addition of additional exotic states with questionable motivation to ensure unification.  So, we will concentrate on the six Higgs doublet possibility.

To achieve successful gauge coupling unification we also need to avoid Landau poles.  The symmetry group $G$ maps Standard Model fermions to (vector-like) partner fermions, but if we need more than three vector-like generations, then $SU(3)_C$ will go non-perturbative before the unification scale.\footnote{With three vector-like generations, $SU(3)_C$ is slightly non-asymptotically free.  This observation also thwarts simple attempts to build models with, e.g., $SO(6)$ symmetry.}  Therefore, $G$ can have two-cycle subgroups that flip between Standard Model and partner fermions, but it cannot have any larger subgroups that cycle between multiple partner fermion copies.

To get a group $G$ of order $6n$ we can use the fact that there are three Standard Model generations.  There can then be a Higgs and a Dark Matter particle for each generation. Formally, there is a three-cycle and a two-cycle subgroup of $G$ for the scalar multiplets:
\be
\mathbf{C}_2 : \quad h_1 \rightarrow \phi_1, \quad h_2 \rightarrow \phi_2, \quad h_3 \rightarrow \phi_3.
\ee
\be
\mathbf{C}_3 : \quad h_1 \rightarrow h_2 \rightarrow h_3 \rightarrow h_1, \quad \phi_1 \rightarrow \phi_2 \rightarrow \phi_3 \rightarrow \phi_1,
\ee
In these kinds of models, all the Yukawa couplings are equal before $G$ is broken, and one can arrange for the $G$-violating interactions to generate both the fermion mass hierarchy through a see-saw mechanism as well as CKM mixing.  

However, if there really is a separate Higgs for every generation, then each Higgs needs to get a vev.  After $G$ is broken, each Yukawa matrix will have the form
\be
\lambda_{\rm eff} v_{\rm eff} = \lambda_1 v_1 + \lambda_2 v_2 + \lambda_3 v_3,
\ee
where $v_i = \langle h_i \rangle$ and $v_{\rm eff}^2 = v_1^2 + v_2^2 + v_3^2$ is the effective electroweak scale.  So instead of having just one source of flavor violation in the matrix $\lambda_{\rm eff}$, there are three sources of flavor violation in the matrices $\lambda_i$.  In general, there will be flavor-changing neutral currents from the tree-level exchange of the extra Higgs multiplets \cite{GlashowWeinberg}.  While it may be possible to fine-tune some of the coefficients of the $\lambda_i$ to make the FCNCs small, generic six doublet models are ruled out by flavor problems.

The danger of FCNCs does not limit us to just two-doublet models, though.  As we show in App.~\ref{sec:approxgauge}, we can arrange for four-doublet or eight-doublet models through repeated uses of two-cycle symmetries.  These models unify much better than the Standard Model, and depending on the sign and magnitude of GUT scale threshold corrections, could conceivably be incorporated in unified models. To avoid additional sources of flavor violation while still having the doublets couple to fermions in irreducible ways, it is necessary to impose \emph{ad hoc} approximate symmetries.  In summary, it is fair to say that gauge coupling unification appears difficult to accommodate without a rather baroque structures.  

Perhaps the simplest possibility to incorporate unification is to adding a single vector-like fermion doublet without a Standard Model partner.  This would unify at with the same precision as a six-Higgs doublet, but it is hard to motivate the presence of the extra fermionic doublet. 

\section{Discussion}

With the first LHC test beam planned for 2007, it is important for model builders to explore many different TeV scale possibilities.  Whether or not environmental selection ultimately explains the electroweak hierarchy, unnatural theories exhibit qualitatively different structures from natural theories.  Examples like the one presented in this paper emphasize that the hierarchy problem is not the only motivation for Standard Model partners, and it is therefore important not only to ascertain the quantum numbers of TeV scale particles at the LHC but also to glean their purpose.   

Others have explored unnatural theories that give the correct Dark Matter relic abundance, often focusing on a minimal particle content \cite{StrumiaDM,NewMinimal}.   These theories often represent an extreme challenge for future colliders. It is encouraging that the simple requirement of natural scalar Dark Matter allows for interesting and well-motivated non-minimalities with robust LHC signatures.  Environmental selection simply offers a justification for why certain parameters can be unnaturally small, but it does not require that every question beyond the Standard Model has an environmental answer.

This model is best motivated by the assumption that the both the cosmological constant and the electroweak scale are chosen by environmental considerations in a landscape of vacua. The smallness of the cosmological constant can be ensured by the ``structure principle'' of Weinberg \cite{WeinbergCC}, while the electroweak vev is set by the ``atomic principle'' \cite{NimaSavas,Donoghue}. Once the electroweak vev is set to the weak scale, it is the dynamics of the theory outlined here that assures us that we get the correct Dark Matter abundance.  This model gives one possible explanation why the Dark Matter abundance is consistent with a weak scale particle:  symmetries relate the weak scale and the Dark Matter scale.    So far, no attempt has been made to assess the frequency with which this type of structure would occur in a String Theory landscape, but given our ignorance about infrared symmetries, it seems plausible that the Higgs and the Dark Matter could end up in the same multiplet.

\section*{Acknowledgments}  We thank N.~Arkani-Hamed, R.~Harnik, R.~Kitano, Y.~Nomura, J.~Wacker, and J.~Wells for illuminating conversations. Thanks to A.~Pukhov for assistance with the \texttt{micrOMEGAs} and \texttt{LanHEP} packages.  The work of AP is supported by the Michigan Center for Theoretical Physics, and the work of JT is supported by a fellowship from the Miller Institute
for Basic Research in Science.

\appendix

\section{Four- and Eight-Doublet Models}
\label{sec:approxgauge}

We can construct four- or eight-doublet models without any flavor problems that unify better than the Standard Model.  The main issue in constructing a natural scalar Dark Matter model with multiple doublets is that the doublets must couple to the Standard Model fermions in an irreducible way.   If there were a linear combination of doublets that did not have some kind of Yukawa coupling, then that linear combination could be given a mass term independent of the Higgs mass term, and the fine-tuning for the Higgs mass would not guarantee that all doublets are light.  We need to start with a (finite) symmetry group $G$ of the same order as the number of doublets, and the Yukawa interactions have to transform non-trivially under $G$.

The trick for creating four- or eight-doublet models is to realize that we can expand the exchange symmetries without needing to introduce any more fermions than in the model from Sec.~\ref{sec:themodel}.  For example, if we have two Higgs doublets $h_1$ and $h_2$, and two Dark Matter doublets $\phi_1$ and $\phi_2$, then we can define the exchange symmetries
\be
\mathbf{S}^A_2 \colon \quad  h_1 \rightarrow \phi_1, \quad h_2 \rightarrow \phi_2, \quad q \leftrightarrow Q, \quad \ell \leftrightarrow L.
\ee
\be
\mathbf{S}^B_2 \colon \quad  h_1 \rightarrow \phi_2, \quad h_2 \rightarrow \phi_1,  \quad u^c \leftrightarrow U^c, \quad d^c \leftrightarrow D^c, \quad e^c \leftrightarrow E^c.
\ee
The up-type Yukawa couplings consistent with these symmetries are:
\be
\mathcal{L}_{\rm Yukawa} = \lambda_u \left(q h_1 u^c + Q \phi_1 u^c + q \phi_2 U^c + Q h_2 U^c \right).
\ee
Note that we have to impose some \emph{ad hoc} symmetries (like a $U(1)$ symmetry that transforms just $U$, $U^c$, $\phi_2$, and $h_2$) to forbid the Yukawa interactions with $1 \leftrightarrow 2$, otherwise the linear combination $h_1 + h_2$ would couple to fermions but the combination $h_1 - h_2$ would not.  Only $h_1$ needs to get a vev, so $h_2$ can be an inert doublet.  In fact, based on arguments similar to \eq{eqn:splitting}, this is what is expected.  If the \emph{ad hoc} symmetries are unbroken at low energies, then both $\phi_1$ and $\phi_2$ are stable and contribute to $\Omega_{\rm DM}$. The $h_{2}$ might decay into the $\phi_{i}$ if kinematically allowed, but in principle might be stable as well.

Building an eight doublet model requires no additional guile; we simply need three different types of exchange symmetries to act on $h_i$ and $\phi_i$.  One possibility is to use the fact that there are both leptons and quarks in the model:
\be
\mathbf{S}^A_2 \colon \quad h_1 \leftrightarrow \phi_1, \quad h_2 \leftrightarrow \phi_2, \quad h_3 \leftrightarrow \phi_3, \quad h_4 \leftrightarrow \phi_4, \quad q \leftrightarrow Q, \quad \ell \leftrightarrow L.
\ee
\be
\mathbf{S}^B_2 \colon \quad h_1 \leftrightarrow \phi_2, \quad h_2 \leftrightarrow \phi_1, \quad h_3 \leftrightarrow \phi_4, \quad h_4 \leftrightarrow \phi_3, \quad u^c (d^c) \leftrightarrow U^c (D^c), \quad \ell \leftrightarrow L.
\ee
\be
\mathbf{S}^C_2 \colon \quad h_1 \leftrightarrow \phi_3, \quad h_2 \leftrightarrow \phi_4, \quad h_3 \leftrightarrow \phi_1, \quad h_4 \leftrightarrow \phi_2, \quad q \leftrightarrow Q, \quad e^c \leftrightarrow E^c.
\ee
The obvious fourth $\mathbf{S}_2$ where you exchange the quark and lepton singlets is equal to the product  $\mathbf{S}^A_2 \times \mathbf {S}^B_2 \times \mathbf{S}^C_2$.  As long as one imposes the appropriate \emph{ad hoc} symmetries, one can use these exchange symmetries to guarantee that every linear combination of $h_i$ or $\phi_i$ couples to at least some fermions.

\end{document}